\documentclass[sigchi, review=false]{acmart}
\usepackage{booktabs} 
\usepackage{balance}

\setcopyright{acmcopyright}

\copyrightyear{2019}
\acmYear{2019}
\setcopyright{acmcopyright}
\acmConference[CHI 2019]{CHI Conference on Human Factors in Computing Systems
Proceedings}{May 4--9, 2019}{Glasgow, Scotland Uk}
\acmBooktitle{CHI Conference on Human Factors in Computing Systems Proceedings
(CHI 2019), May 4--9, 2019, Glasgow, Scotland Uk}
\acmPrice{15.00}
\acmDOI{10.1145/3290605.3300376}
\acmISBN{978-1-4503-5970-2/19/05}

\begin{document}
\newcommand{\papername}{SottoVoce}
\title[SottoVoce: An Ultrasound Imaging-Based Silent Speech Interaction]{SottoVoce: An Ultrasound Imaging-Based Silent Speech Interaction Using Deep Neural Networks}

\newcommand{\commentout}[1]{}

\commentout{
\author{Naoki Kimura}
\affiliation{%
  \institution{The University of Tokyo}
  \city{Tokyo}
  \state{Japan}
}
\email{kimura-naoki@g.ecc.u-tokyo.ac.jp}

\author{Michinari Kono}
\affiliation{%
  \institution{The University of Tokyo}
  \city{Tokyo}
  \state{Japan}
}
\email{mchkono@acm.org}

}

\author{Naoki Kimura}
\affiliation{%
  \institution{The University of Tokyo}
  \city{7-3-1, Hongo, Bunkyo-ku}
  \state{Tokyo}
  \postcode{113-0033}
  \country{Japan}
}
\email{kimura-naoki@g.ecc.u-tokyo.ac.jp}

\author{Michinari Kono}
\affiliation{%
  \institution{The University of Tokyo}
  \city{7-3-1, Hongo, Bunkyo-ku}
  \state{Tokyo}
  \postcode{113-0033}
  \country{Japan}
}
\email{mchkono@acm.org}

\author{Jun Rekimoto}
\authornote{corresponding author}
\orcid{0000-0002-3629-2514}
\affiliation{%
 \institution{The University of Tokyo}
\city{7-3-1, Hongo, Bunkyo-ku}
\state{Tokyo}
\country{Japan}
\postcode{113-0033}
}
\affiliation{%
 \institution{Sony Computer Science Laboratories}
\city{3-14-13 Higashigotanda, Shinagawa-ku}
\state{Tokyo}
\country{Japan}
\postcode{141-0022}
}
\email{rekimoto@acm.org}

\renewcommand{\shortauthors}{Kimura, Kono and Rekimoto}

\begin{abstract}
The availability of digital devices operated by voice is expanding rapidly. However, the applications of voice interfaces are still restricted. For example, speaking in public places becomes an annoyance to the surrounding people, and secret information should not be uttered. Environmental noise may reduce the accuracy of speech recognition. To address these limitations, a system to detect a user's unvoiced utterance is proposed. From internal information observed by an ultrasonic imaging sensor attached to the underside of the jaw, our proposed system recognizes the utterance contents without the user's uttering voice. Our proposed deep neural network model is used to obtain acoustic features from a sequence of ultrasound images. We confirmed that audio signals generated by our system can control the existing smart speakers. We also observed that a user can adjust their oral movement to learn and improve the accuracy of their voice recognition. 
\end{abstract}

%
%



\begin{teaserfigure}
  \includegraphics[width=\textwidth]{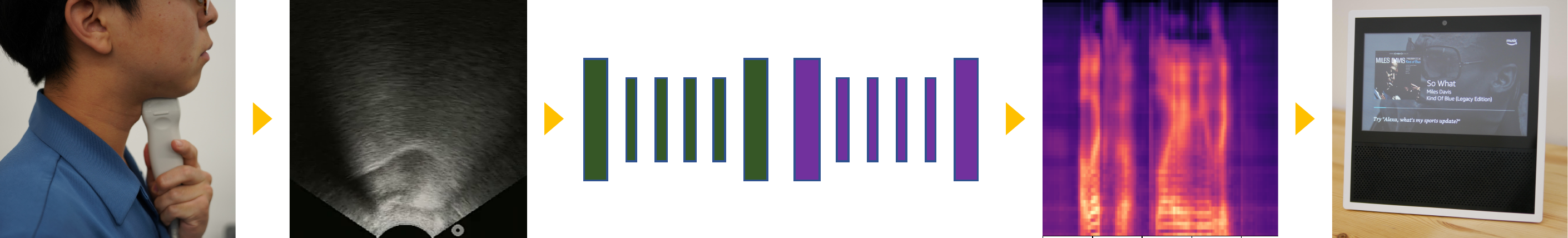}
  \caption{SottoVoce silent voice system: an ultrasonic echo probe attached under the jaw that reads the internal situation while the user is speaking without actually emitting a voice. By recognizing ultrasound images using deep convolutional neural networks, the user's voice is resynthesized and can be used to control the existing speech interaction systems such as smart speakers.}
  \label{fig:teaser}
\end{teaserfigure}

\maketitle

\section{Introduction}

Smart devices that are controllable by speech (i.e., speech interaction) are being used in many situations. Smartphones, smart speakers, car-navigation systems, and various home appliances can be controlled by speech and have presented many interactive potentials~\cite{Rudnicky:1989:DVI:100964.100972}. Speech interaction does not require visual attention, and it can be used in a dark environment such as a bedroom. Owing to the recent progress of speech-recognition technology and the naturalness of speech synthesis, speech interaction is becoming an irreplaceable process in human--computer interaction. Speech can also be used while the user of a speech-interaction device is performing other tasks, such as driving, cooking, homeworking, or using a traditional personal computer. For example, while the user is concentrating on a computer-screen and interaction devices such as a keyboard and a mouse, they can still operate other devices with voice interaction. 

However, as for the speech interface, two challenges must be overcome. First, using the interface in public places presents limitations. In addition to being an annoyance to the surrounding people, disclosing personal information or secret information by uttering it in public is risky in terms of information security. Next, it cannot be used in a noisy environment, because the accuracy of speech recognition may be declined. These issues are particularly acute when trying to use a speech interface to interact with wearables or mobile computers.

To overcome these challenges, research on ``silent-speech recognition'' has been conducted~\cite{Denby:2010:SSI:1746726.1746804}. For example, by applying a method known as lip reading, images of the mouth of the speaker or the entire face are captured by a camera, and the content of the utterance is estimated from those images~\cite{DBLP:journals/corr/WandKS16}. If the user could simply {\em mouth} the utterance without actually voicing it, it would be possible to use such a voice interaction in public places. However, with the camera method, it is necessary to install a camera in front of the face, and its form factor renders it unsuitable for wearables or mobile applications. Other approaches of the studies on ``non-audible murmur'' (NAM)~\cite{HIRAHARA2010301,1200069} attempts to recognize utterances with a microphone or an accelerometer worn on the skin or throat of the user. In this case, the user speaks with articulated respiratory sound without vocal-fold vibration (namely, whispering). However, to be  recognized accurately by the system, a user's whispering voice tends to be noticeable by other people nearby. Furthermore, some studies attempt to estimate speech by estimating the movement of muscles near the oral cavity by electromyography (EMG)~\cite{10.1007/978-3-642-14097-6_96, Kapur:2018:APW:3172944.3172977}. However, the estimation of free utterances with EMG is still difficult; instead, it is a type of gesture recognition using the movement of the oral cavity. Thus, the number of detectable commands is limited, and the user has to learn new gesture skills instead of using their existing speaking skills.

Instead of using the above-described approaches, we focus on using ultrasonic imaging~\cite{acousticMicroscopy}. Ultrasonic-imaging technology recognizes the internal status in the body by measuring the reflection time of ultrasonic waves radiated into the body. This technology is widely used to grasp the condition of the internal organs for medical purposes. In recent years, small and lightweight systems that can be directly connected to smartphones (e.g., Vscan Extend, General Electronic Company) have appeared. If it were possible to attach a small ultrasonic imaging head around the neck to sense the situation in the oral cavity and convert it to acoustic information, it would be a useful device for communicating with speech-capable devices without actually speaking aloud; namely, ``silent voice interaction'' would be possible.

Silent voice interaction by ultrasonic imaging demonstrates two potential advantages over other approaches. First, the ultrasonic-imaging head can be miniaturized, and it can constitute a device with an inconspicuous shape such as a collar. It would be an important feature for designing wearable silent-voice systems. Next, by recognizing the situation in the oral cavity, it would be possible to measure the movement of the tongue, which cannot be observed from the outside. It would thus be possible to reproduce sound more accurately.

Studies have been conducted on silent speech using ultrasound imaging, but many of them are used in combinations with 
lip or face images; therefore, a camera must be placed in front of the user~\cite{Hueber:2010:DSS:1746726.1746805}. This configuration presents a limitation when it is used as a wearable interface device. Recent researchers have challenged to use deep neural networks with ultrasound imaging for silent speech~\cite{Csap2017DNNBasedUC,dnnultrasound18}; however, they are not based on convolutional neural networks and are not validated with data retrieved from a mouth movement without speaking. They are only validated with mouth movement when a user is actually emitting a voice. We present a significant step forward using convolutional neural networks and proof-of-concept validations via actual silent speech, interacting with an unchanged smart speaker (Amazon Alexa).

Herein, a silent-voice interaction system-called ``SottoVoce'' based only on ultrasonic images is described (Figure~\ref{fig:teaser}). By combining two types of deep neural networks, this system can be trained to generate voice signals from a sequence of images captured from an ultrasonic-imaging device. Our contributions can be summarized as the following two topics:
 
\begin{itemize}
\item A two-level model of deep convolutional neural networks to convert ultrasonic images to actual sounds is proposed.
\item As a proof of concept, a silent-voice system was developed, and it was shown that the system could control a voice-controlled device (in this case, Amazon Echo) without modifications.

\end{itemize}

\section{Related Work}

\subsection{Silent Speech}
Silent-speech interfaces have been studied using various technologies and methods~\cite{Denby:2010:SSI:1746726.1746804,Ji:2018:USS:3198931.3199070}. Lip reading~\cite{DBLP:journals/corr/WandKS16} or facial images~\cite{ephrat2017improved} can be used to estimate speech uttered by a subject without using audio information.

SilentVoice~\cite{Fukumoto:2018} is an {\it ingressive speech} approach that captures extremely soft speech. Electromagnetic articulography (EMA) has been used to develop brain--computer interfaces~\cite{10.1371/journal.pcbi.1005119} and other interactive applications~\cite{Wang2014}. Magnets can be attached to the subject to detect silent speech~\cite{Gonzalez:2016:SSS:2936017.2936059,Hofe:2013:SSR:2400751.2401095,FAGAN2008419}. Electroencephalogram (EEG)~\cite{Porbadnigk2009EEGbasedSR} and Electromyography  (EMG)~\cite{10.1007/978-3-642-14097-6_96,1566521} are also typical methods for silent-speech recognition. Particularly, EMG has been applied for interactive purposes and applications concerning human--computer interaction (HCI)~\cite{Manabe:2003:USR:765891.765996}; for example, controlling a web browser~\cite{Jorgensen:2005:WBC:1042440.1043301}. A recent example proposed by Kapur et al.~\cite{Kapur:2018:APW:3172944.3172977} used multiple electrodes to sense neuromuscular signals of a subject in an internal speech. In addition to the methods introduced above, other human--facial electrical potentials have been combined and measured ~\cite{HIRAHARA2010301,Sahni:2014:TEI:2634317.2634322}.

Ultrasound imaging has also been used for silent-speech recognition in the field of speech processing \cite{1326078,Hueber2011StatisticalMB}. In one study, which focused on singing, sung vowels could be synthesized based on the ultrasound and video of the lips~\cite{jaumardhakoun:hal-01529630}. In a similar study, a silent-speech interface using ultrasound and optical images of the tongue and lips was developed~\cite{Hueber:2010:DSS:1746726.1746805,Hueber2008}. In another study, a mapping technique for automatically generating animations of the tongue movement from raw ultrasound images was created~\cite{FABRE201763}. 

Other approaches combine the techniques described above with deep neural networks, such as BCI applications~\cite{Bocquelet2014}, myoelectric signals~\cite{7280404}, and 
lip reading with long short-term memory (LSTM)~\cite{DBLP:journals/corr/WandKS16}. Combining deep neural networks and ultrasound imaging for silent-speech interfaces has also been considered. For example, ultrasound imaging was used to capture the tongue movement with such a combination with deep neural networks~\cite{Csap2017DNNBasedUC}. Moreover, the fundamental frequency (F0) curve, which had been considered unpredictable, was estimated using a deep neural network~\cite{Grosz2018}. In addition, it has been suggested that a global and visuo--acoustic modeling approach called ``Eigentongues'' performs better than tongue-contour modeling when using neural networks~\cite{4217312}. A new benchmark for silent-speech research based on deep neural networks has also been proposed~\cite{Ji:2018:USS:3198931.3199070}. These approaches are still at the basic proof-of-concept level, and none have been evaluated in terms of controlling the existing speech-interaction appliances such as smart speakers.

Following the prior approaches described above, we aim to develop a system that allows silent speech to interact with voice-controlled devices and interactive purposes in a HCI context. Furthermore, to improve the performance of the developed system, we aim for the interaction between human and artificial neural networks for performance improvement.

\subsection{Non-Auditory Inputs and Interaction}
EarFieldSensing~\cite{Matthies:2017:ENI:3025453.3025692} is a gesture-recognition technology based on electric-field sensing. CanalSense~\cite{Ando:2017:CFM:3126594.3126649} senses changes in air pressure in the ear canals that occur when the face is moved. Tongue-in-Cheek~\cite{Goel:2015:TUW:2702123.2702591} senses the movement of the tongue using the X-band Doppler radar for facial--gesture recognition. Other methods that use EMG~\cite{Zhang:2014:NTM:2556288.2556981} or combinations of brain and muscle signal sensing~\cite{Nguyen:2018:TYT:3210240.3210322} are also notable.

The techniques above have been utilized for arm-gesture recognition; however, an approach using ultrasound imaging for that purpose has been demonstrated, i.e., EchoFlex~\cite{McIntosh:2017:EHG:3025453.3025807}. It is an interaction sensor that recognizes movements of the forearm muscle using ultrasound imaging. The results of that study indicated that the sensor performed well and can potentially be supplanted to other prior approaches. They also inspired the authors to consider using ultrasound imaging for silent-voice interaction using the inner part of the mouth.

\begin{figure*}
\includegraphics[width=0.9\textwidth]{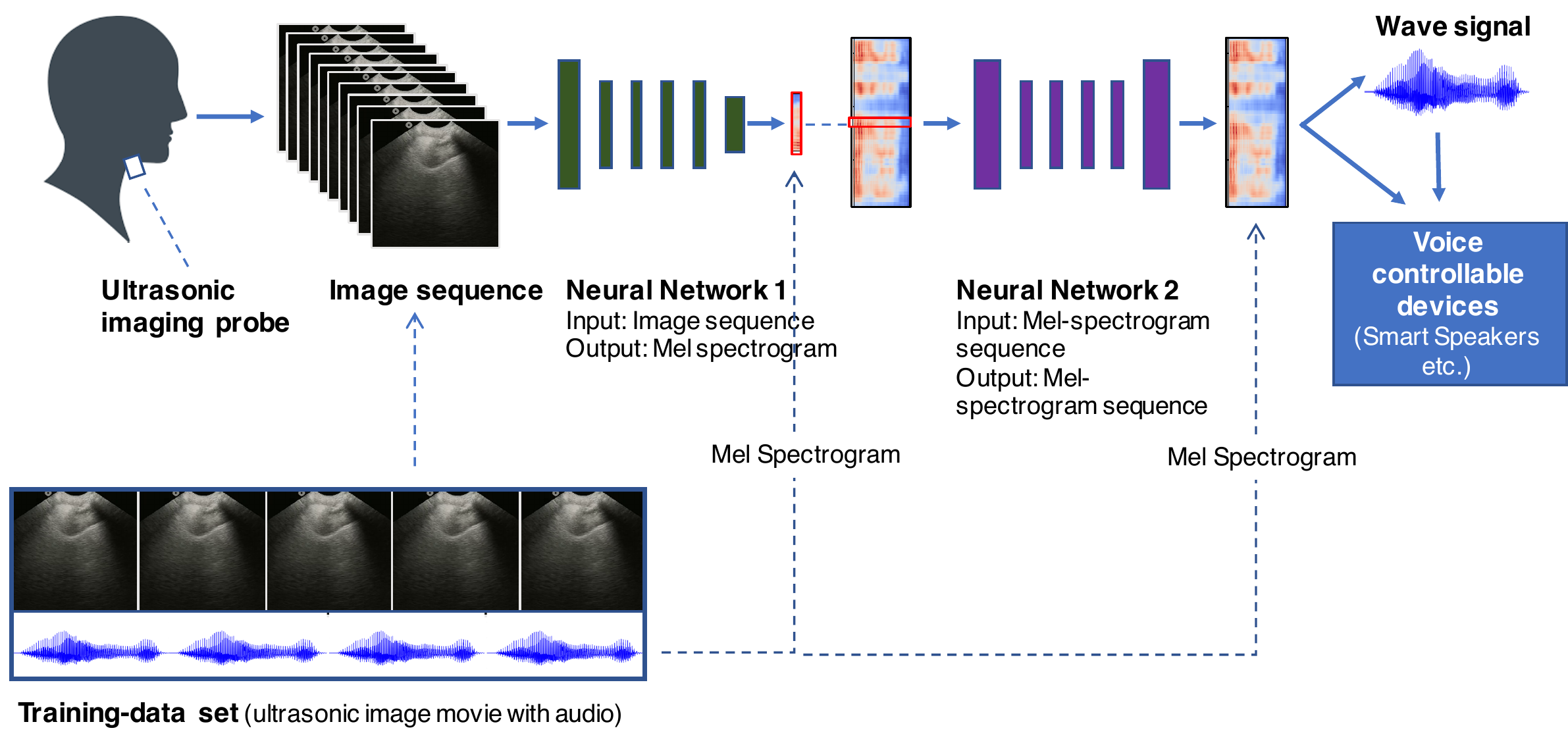}
\caption{SottoVoce system overview}
\label{fig:system}
\end{figure*}

\subsection{Interacting with Smart Devices}
Owing to the development of mobile and smart devices, we can now interact with them frequently through various methods~\cite{Klopfenstein:2017:RBS:3064663.3064672}. We may use our voice as an input method~\cite{Rudnicky:1989:DVI:100964.100972}, and studies to overcome issues concerning such voice-controlled user interfaces have been presented~\cite{Myers:2018:PUO:3173574.3173580}. One typical interface (a personal ``agent'' hereafter) is Amazon Echo, typically known as ``Alexa,'' with which people can communicate and chat. The effect of this agent on people has been researched extensively, and the results show that it is an effective agent for satisfying or influencing our lives~\cite{Kocielnik:2018:DWR:3196709.3196784,Purington:2017:AMN:3027063.3053246,Sciuto:2018:HAW:3196709.3196772}.

The role of artificial intelligence (AI) has become more important for tasks other than personal agents like Alexa, and we now interact and collaborate with AIs; for example, when sending text messages~\cite{Hohenstein:2018:AMI:3170427.3188487} and designing objects~\cite{Fiebrink2017}. The hybrid existence and interaction of humans and AI is an impressive topic, which can overcome social issues and improve the quality of our lives. Glove Talk II~\cite{Fels:1995:GAG:223904.223966}, which was presented in 1995, is a gesture-to-speech system that translates hand gestures into 10 control parameters of a speech synthesizer using neural networks. However, to use the system, the user requires a long-term training of approximately 100 hours. In the present study, referring to this work, we apply the recently improved technologies of deep-neural networks to create the hybrid interaction of humans and AI, where the user (a human) learns to adapt and utilize the AI system to achieve a better interaction.

\section{System Architecture of SottoVoce}

The architecture of the proposed system for generating sound from ultrasonic images is shown in Figure~\ref{fig:system}. In general, the goal of the system is to transfer certain sequence representations (in this case, ultrasonic images) into other sequence representations (in this case, speech). This goal is similar to that of text-to-speech systems~\cite{DBLP:journals/corr/WangSSWWJYXCBLA17}, voice-transfer systems~\cite{deepvoice}, and lip reading or face-to-voice systems~\cite{ephrat2017improved}. Inspired by these systems, the proposed system uses two neural networks. 

The first neural network (`Network~1' in Figure~\ref{fig:system}) transfers a time series of ultrasonic images to a sound-representation vector. We used a 64-dimensional Mel-scale spectrum with the frequency range of $300\:Hz$ to $8,000\:Hz$, sampled every $20\:ms$, as a sound-representation vector. Subsequently, the translated sound representations constitute a series of sound-representation vectors (i.e., a spectrogram). This sound spectrogram can be converted to an audio signal. In addition, to refine the quality of those vectors, they are also transferred by the next neural network (``Network~2'' in Figure~\ref{fig:system}), which generates a series of sound-representation vectors of the same length as that of the input sound-representation vectors. Finally, the output vectors are converted to an actual audio signal.

These two networks are speaker dependent; accordingly, to train them, the system requires a set of ultrasonic-imaging videos captured while the user speaks various speech commands.

\subsection{Ultrasonic Imaging Device}

\begin{figure}
\centerline{
\hbox{
\includegraphics[width=0.22\textwidth]{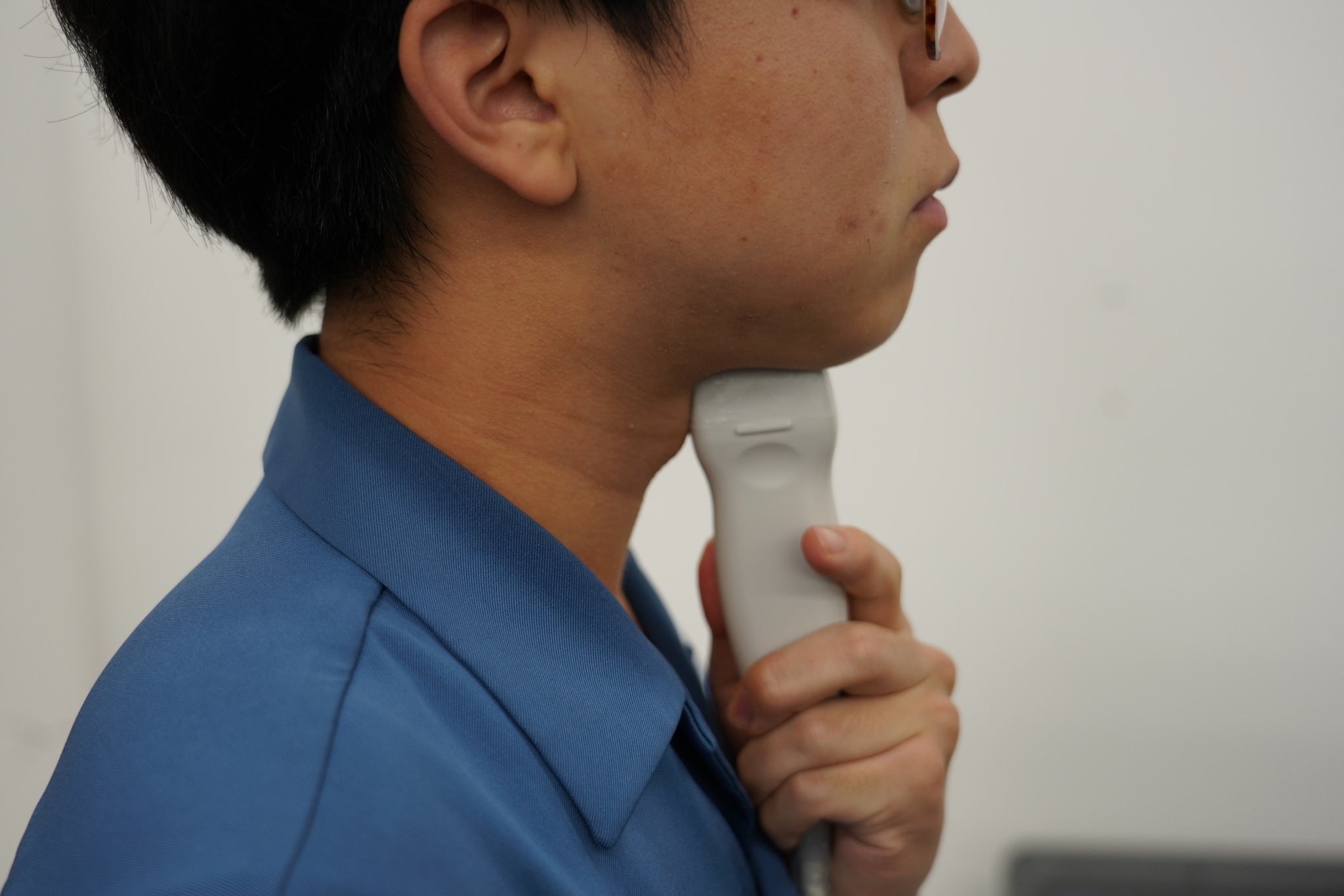}
\includegraphics[width=0.22\textwidth]{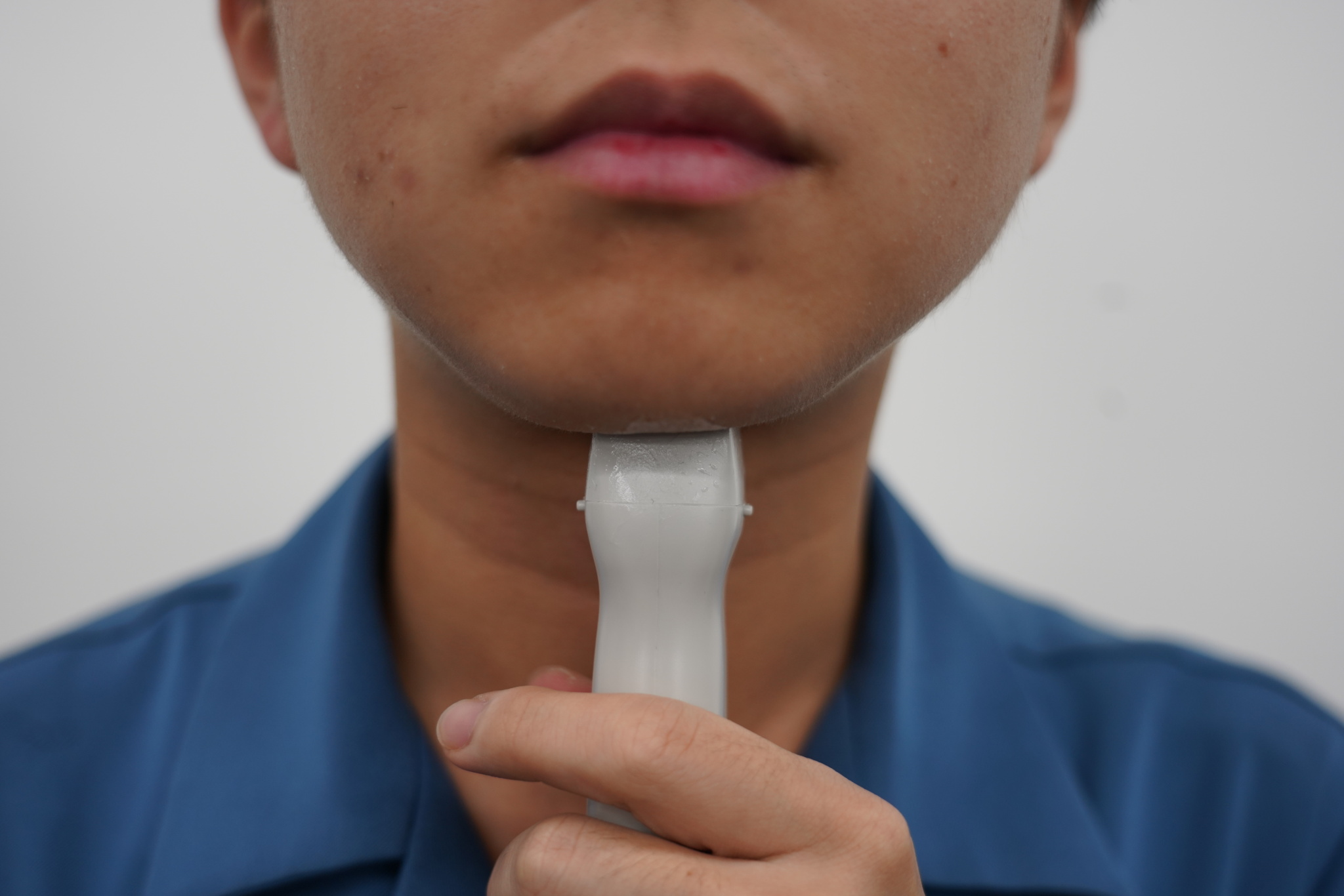}
}
}
\caption{Ultrasonic imaging probes}
\label{fig:probe} 
\vspace{-0.4cm}
\end{figure}

\begin{figure}
\includegraphics[width=0.3\textwidth]{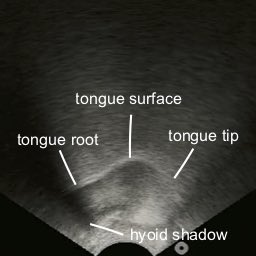}
\caption{Obtained ultrasonic image from probes attached to the jaw from underneath.}
\label{fig:ultra}
  \vspace{-0.4cm}
\end{figure}

The CONTEC CMS600P2 Full Digital B-Ultrasound Diagnostic System was used as the ultrasonic-imaging device. A user attaches a 3.5-MHz convex-type ultrasonic imaging probe under the jaw (Figure~\ref{fig:probe}). This system provides a screen output port to be connected to the display monitor. In addition, a display-digitizing unit was used for converting the signal sent to the display to an MPEG-4 movie file. Figure~\ref{fig:ultra} shows the obtained ultrasonic image.

We found a delay of ultrasonic images and sound capturing. To compensate for this delay, we examined the corresponding utterance and tongue movement in the video and estimated a delay of $300\: ms$. We subsequently adjusted to this delay in the training data.

\begin{figure*}
\includegraphics[width=0.95\textwidth]{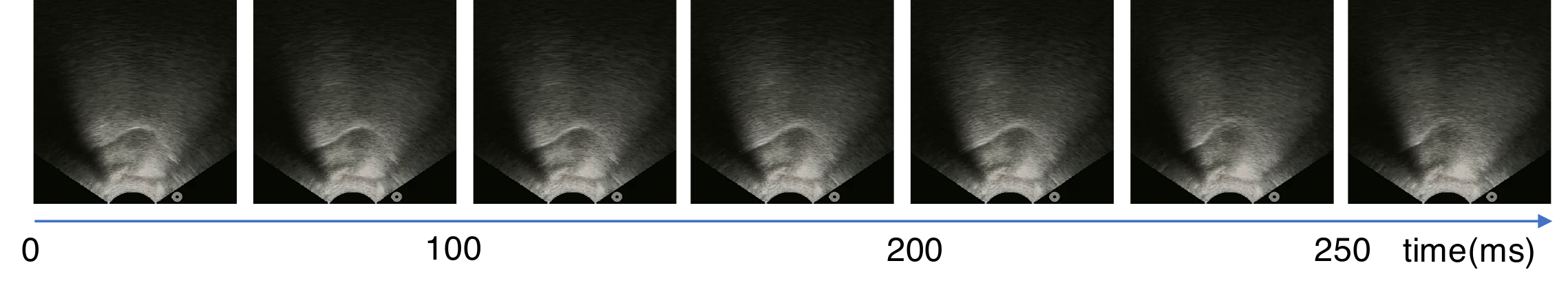}
\caption{A series of ultrasonic images of the throat of a subject about to pronounce ``Alexa.''}
\label{fig:ultraimage}
\end{figure*}

\subsection{Network 1}
\begin{figure}
\includegraphics[width=0.4\textwidth]{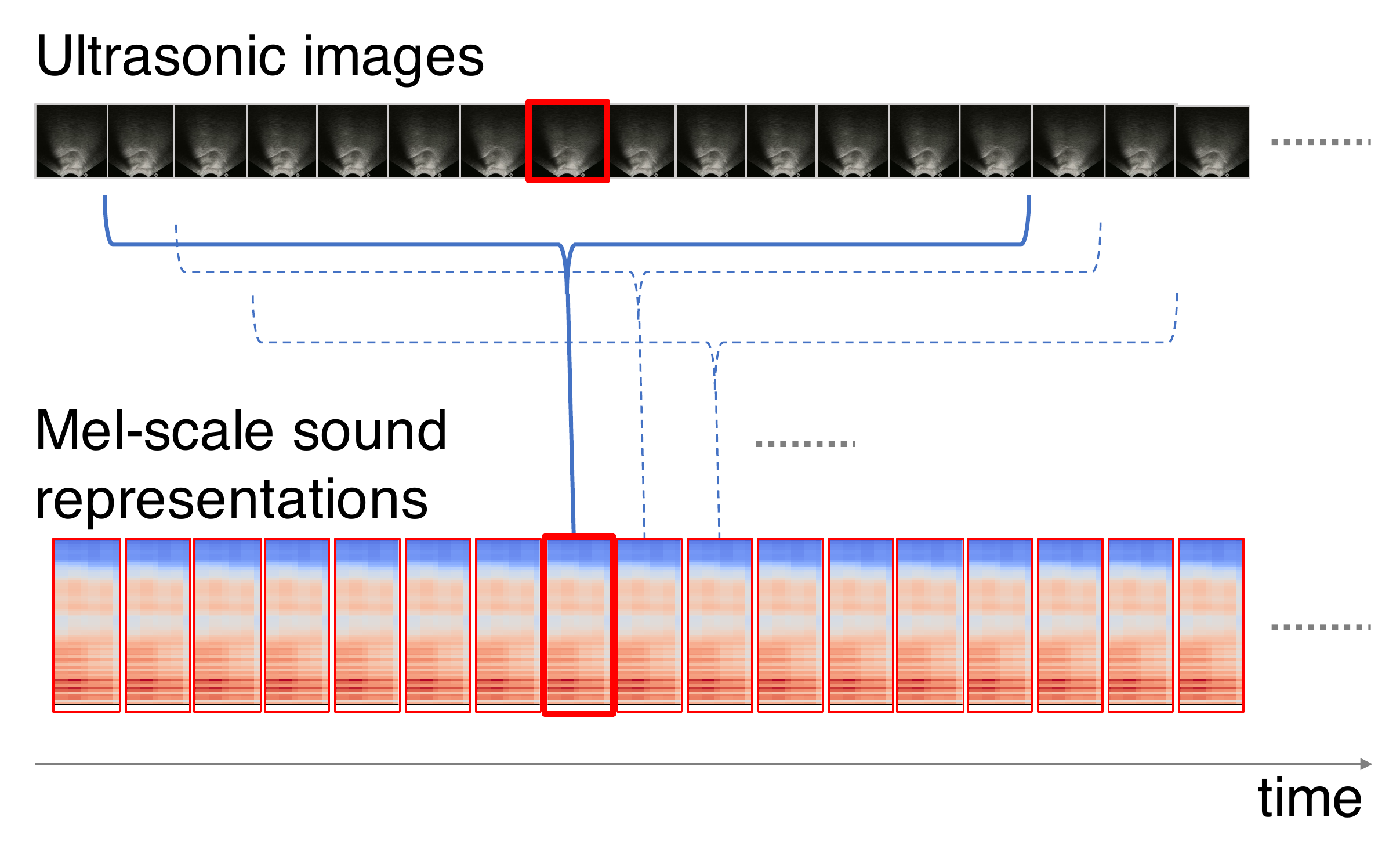}
\caption{Representation of training data. K-sized ultrasonic images are paired with the corresponding sounds (Mel-scaled sound vectors).}
\label{fig:hop}
\end{figure}

Network~1 uses a series of $K$ ultrasonic images (size of $128 \times 128$, monochrome) as the input and generates an $n$-dimensional sound representation (Mel-scale spectrum) as the output. Currently, $K$ of 13 and $n$ of 64 are used. Because the frame rate of the ultrasonic images is 30 frames per second, the duration of $K$ ultrasonic images is thus 400 ms. This time duration covers the static and motion features of the utterance. Samples of the ultrasonic images are shown in Figure~\ref{fig:ultraimage}. The K-size image sequence is prepared repeatedly such that one Mel-scale spectrum is created every 20 ms (i.e., the hop size is 20 ms) (Figure~\ref{fig:hop}).

A sound-representation vector corresponding to the time position at the center of each ultrasonic-image sequence is extracted from the audio signal data, and Network~1 is trained to generate it.

Network~1 is based on a convolutional neural network (CNN). It comprises four layers: {\it Conv2D - LeakyReLU - Dropout - BatchNormalization}, followed by six layers: {\it Flatten - Dense - LeakyReLU - Dropout - Dense - LeakyReLU}. The output size of Network~1 is the same as the length of the sound-representation vector (i.e., 64). Both input images and output vectors are normalized to 0 to one, respectively. The loss function is the mean-squared error, and the optimizer is Adam~\cite{DBLP:journals/corr/KingmaB14}.

\subsection{Network 2}

\begin{figure}
\includegraphics[width=0.48\textwidth]{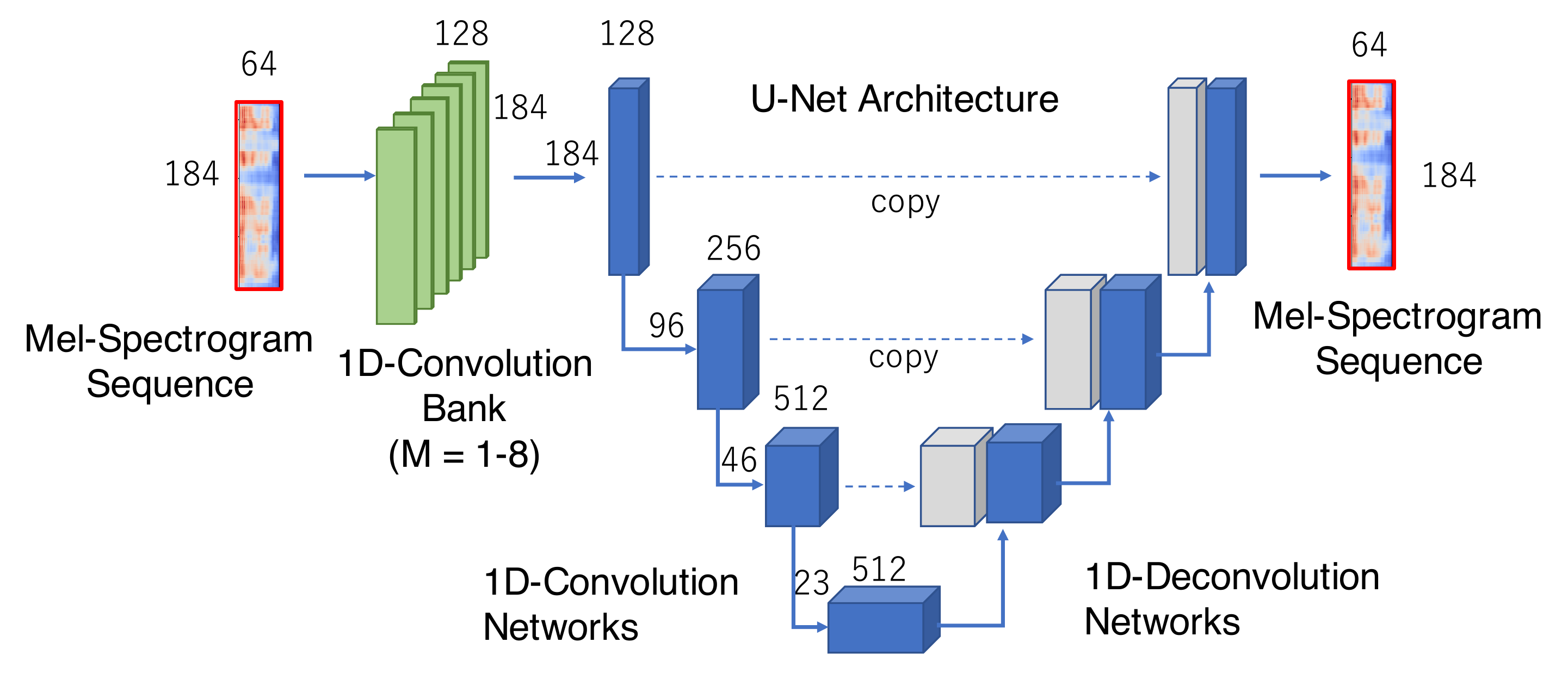}
\caption{Network~2 improves the quality of generated Mel-scaled spectrum sequences. (Note: for the consistency with the illustration of the neural networks, the time axis of the audio-feature vector is shown as the vertical axis.)}
\label{fig:net2}
\end{figure}

To improve the sound quality, Network~2 uses a sequence of sound-representation vectors and generates a sequence of sound-representation vectors with the same length as the input. This model comprises a bank of one-dimensional (1-D) convolutional filters ({\it Conv1D}), with a kernel size from 1 to $M$ ($ M = 8$ is currently used),  followed by the U-Network~\cite{DBLP:journals/corr/RonnebergerFB15} with the three layers of {\it Conv1D - MaxPooling (strides=2) - LeakyReLU - Dropout}, and thrice of {\it DeConv1D - Concatenate} (Figure~\ref{fig:net2}). The 1-D convolutional bank explicitly models the local and contextual information of the input sequence. The following U-Network also improves the quality of the audio sequence with precise localization. Finally, the network generates Mel-scale spectrum vectors.

To train Network~2, Network~1 was used to create Mel-scale spectrum vectors from the images of a training ultrasonic video clip as the input, and the same length Mel-scale spectrum vectors from the audio of the same training video clip as the output. Similarly, in the case of Network~1, the mean-squared error was used as a loss function, and Adam was used as an optimizer.

For simplicity, the time durations of the input and output were fixed to the same value (currently, $3.68\: s$ is used). This duration encompasses many typical speech commands.

\subsection{Sound generation}
Following the neural-networks processing, a sequence of Mel-scale-spectrum sound-representation vectors is converted to an audio signal using the Griffin Lim algorithm~\cite{1164317}. This conversion is possible from the output of Network~1 or the output of Network~2. 

For testing, the generated audio signals are transmitted from an audio speaker, and they can be used to control nearby sound-controlled devices such as a smart speaker. We are also considering taking the audio-waveform signal directly as the audio input information of the speech-controllable device without actually reproducing it as a sound wave.

\subsection{Training}

\begin{figure}
\includegraphics[width=0.45\textwidth]{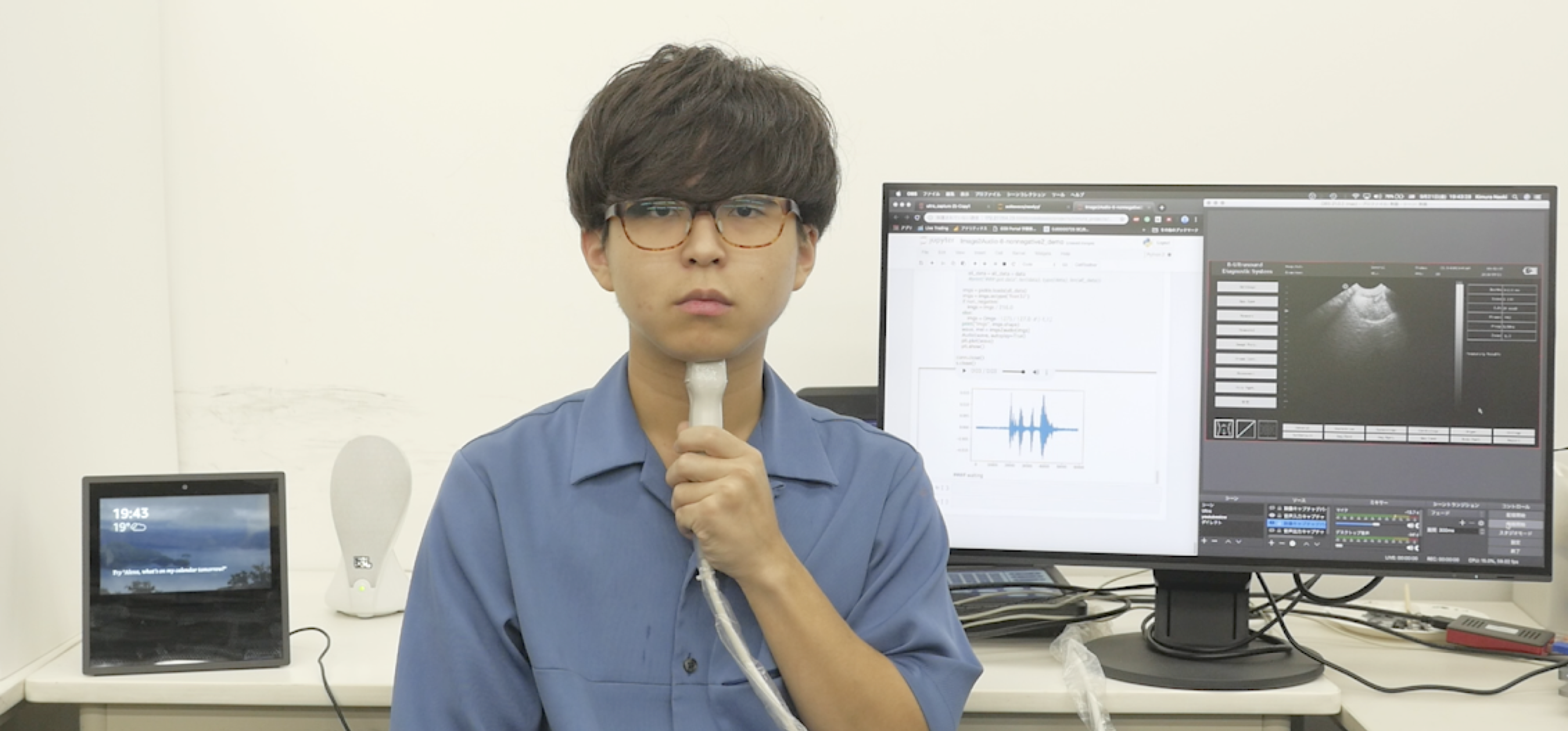}
\caption{Apparatus used for training and evaluating.}
\label{fig:training}
\end{figure}

As for preparing the training data, two collaborators, (28-year old male and 24-year old male) were attached with an ultrasonic imaging probe under their jaws, and were instructed to utter various speech commands. Approximately 500 speech commands were collected from each collaborator (Table~\ref{fig:commands}). For each command, as well as the voice utterance, a video of the ultrasonic images was recorded. The training session was approved by the research ethics committee of the author's institution.

The recorded video was used to train Network~1. The ultrasonic images were rescaled to $128 \times 128$ and used as inputs. The corresponding utterance voice was converted to a Mel-scale spectrum and used as outputs.

The number of test sets for Network~2 was the same as the number of trained video files (approximately 500). To increase the number of test sets, data augmentation by applying Gaussian noise to the input Mel-scale spectrum vectors was used.

As our model is speaker dependent, both Network~1 and Network~2 are  trained for each speaker. Network~1 is trained first and subsequently used to create the dataset for training Network~2.

\subsection{Implementation Details}
The above-described network models were implemented based on the Keras~\cite{chollet2015keras} deep-learning platform with Tensorflow~\cite{tensorflow2015-whitepaper} as the backend, and an NVIDIA GeForce 1080ti as the GPU board. Training Network~1 with 500 speech commands (which creates 35,000 training data pairs for Network~1) required approximately 4 h. Training Network~2 required less than an hour.
comment
As the ultrasonic-imaging device cannot be connected directly to the Ubuntu machine that runs the neural networks, a simple server--client program was developed. It connects the computer that controls the ultrasonic imaging device to the computer that operates the neural networks. The generated audio signals are sent back to the computer with the ultrasonic-imaging device. 

To process ultrasonic images at/with duration/intervals of $3.68\: s$, $2.36 \:s$ is required for the neural networks to process the images. The total processing time (including video processing, neural networks processing, and conversion of the Mel-scale spectrum to an audio wave) was $2.61\:s$.

\begin{figure*}
\centerline{
\includegraphics[width=0.9\textwidth]{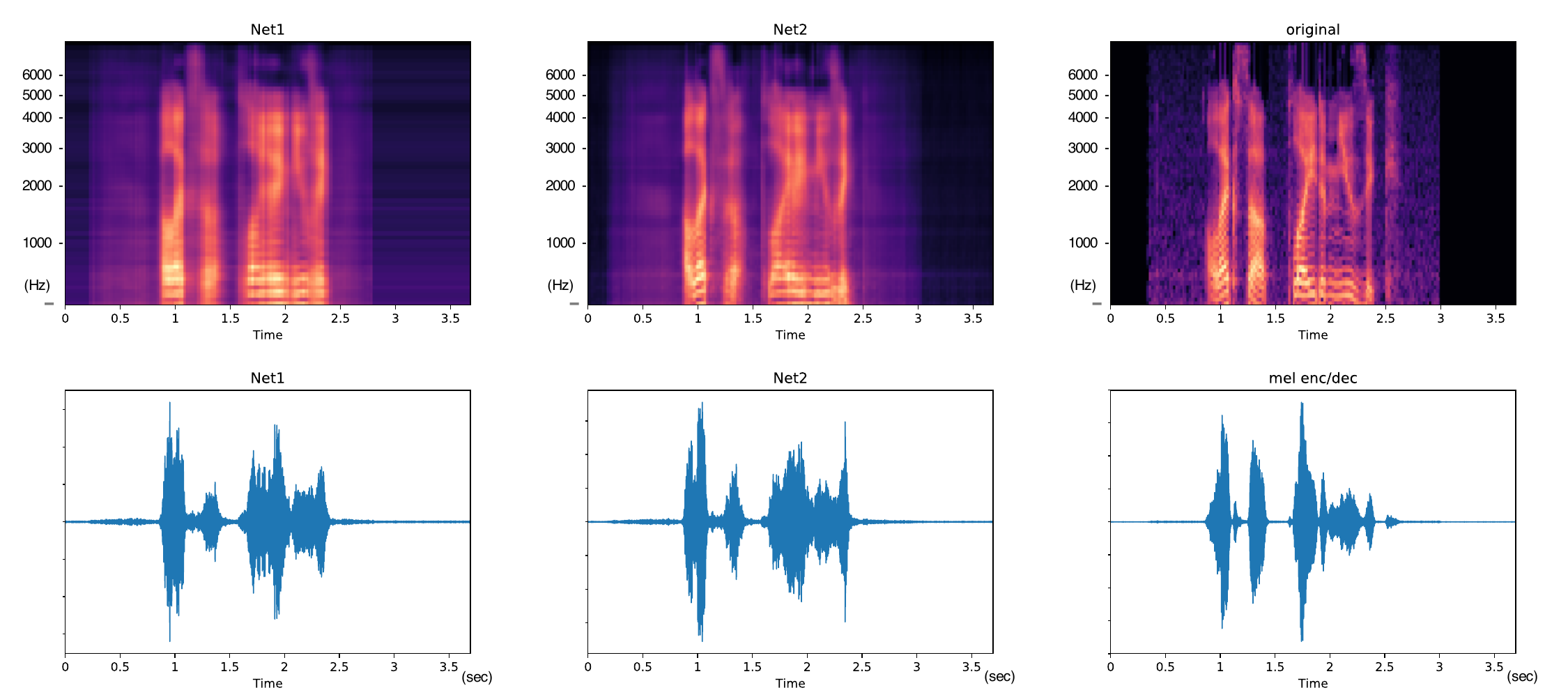} 
}
\caption{Results of the training (shown as waves and Mel-scale spectrogram): Net1: Network~1 results; Net2: Network~2 results; original: original voice encoded by Mel-scale spectrogram and decoded to an audio signal. This is the ``ground-truth'' of this training.}
\label{fig:mel} 
\vspace{-0.2cm}
\end{figure*}

\begin{figure*}
\centerline{
  \includegraphics[width=0.9\textwidth]{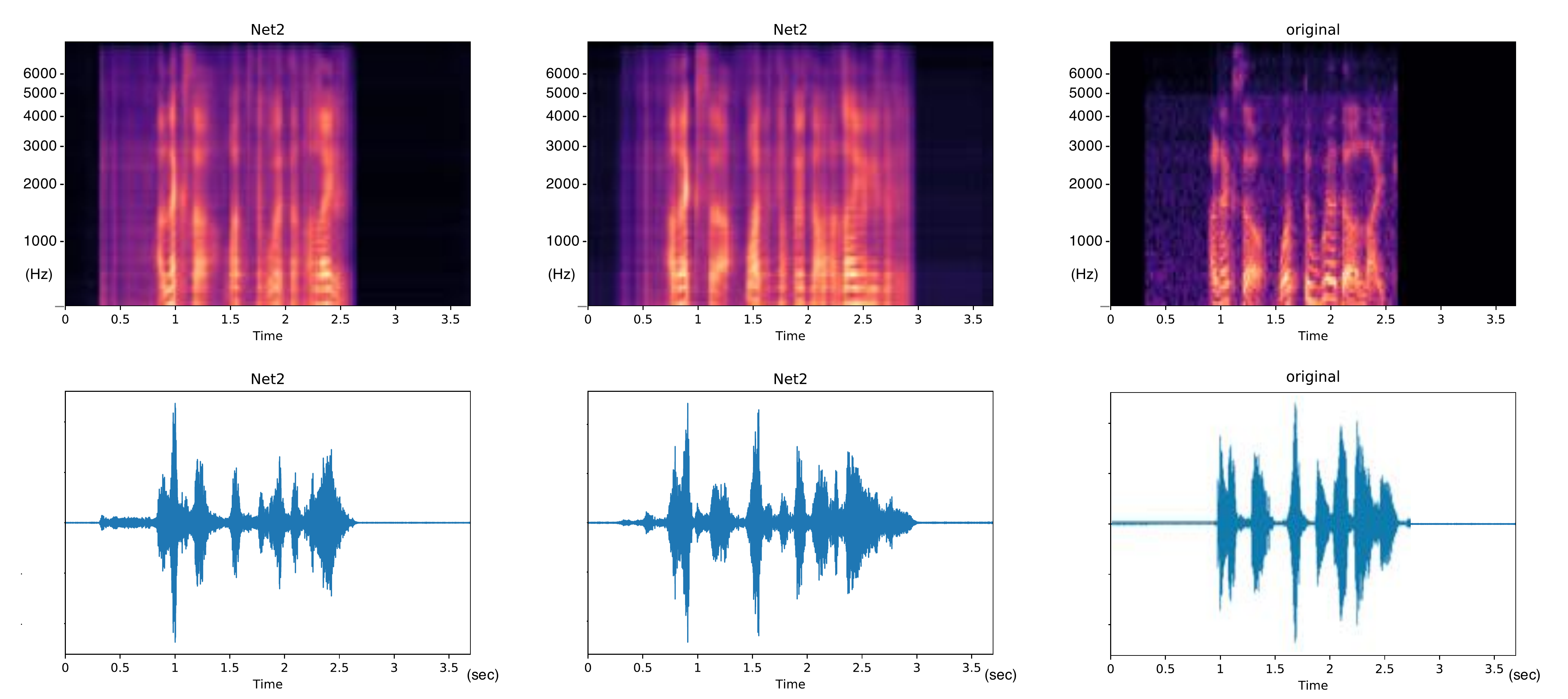}
}
\caption{Comparison of Generated Sounds. Left: Network~2 results from ultrasonic images while a user is emitting a voice; Middle: Network~2 results from ultrasonic images without emitting a voice; Right: the original (Mel-scale spectrogram encoded/decoded) voice}
\label{fig:comparison}
  \vspace{-0.2cm}
\end{figure*}

\section{Results}

The results of converting the ultrasonic images to sound are shown in Figure~\ref{fig:mel}. In the figure, the top-row graphs show the sound-representation vectors (a Mel-scale spectrogram), and the bottom-row graphs are the corresponding waveforms. The graph labeled Net1 is the result of Network~1, that labeled Net2 is the result of Network~1 + Network~2, and that labeled ``original'' is the wave data encoded as a Mel-scale spectrogram and decoded back to the waveform. Thus, the last one is regarded as the ground-truth of the training. Although the difference between the outputs of Network~1 and Network~2 was unclear, we observed that the sound generated by Network~2 was better than that generated by Network~1 (Examples of the output audio signals are given in the supplemental video.)

It is noteworthy that both Network~1 and Network~2 generate natural intonation, which is typically considered to be caused by vocal-fold vibration, not by the internal situation. This result might suggest that the neural networks learn the context of the speech.

The generated sounds emitted from the computer's speaker were subsequently tested with an existing (unchanged) smart speaker (Amazon Echo and Amazon Echo Show), and this test confirmed that the generated sounds can control smart speakers. The speech commands used for training and testing were typical Amazon Alexa commands. For this test, the participants spoke the following four commands, five times each (20 utterances in total): ``Alexa, play music,'' ``Alexa, what's the weather like,'' ``Alexa, what time is it,'' and ``Alexa, play jazz.''

Table~\ref{fig:table} lists the recognition success ratio of Network~1, Network~1 + Network~2, and the original (Mel-scale encoded and decoded) as a ground truth. We confirmed that the combination of Network~1 and Network~2 improves the recognition rate. We also noticed that the trigger word (``Alexa'') is always regenerated clearly. This may because that word is simply the most pronounced word in the training set. 

We also performed the word error rate (WER) measure test using Google's cloud speech-to-text engine~\cite{googlespeech} and the same environment used for the recognition measurement of smart speakers. The WER was 20.61\%, 41.03\%, and 33.56\% for GT, Network~1, and Network~2, respectively (mean of a total of 40 speech commands from two users from Table~\ref{fig:table}). We believe that this will also serve as evidence for the effectiveness of Network~2. Through these studies, we found that commands such as ``what's the weather like?'' had a high recognition rate for all conditions, while commands with shorter terms such as ``play jazz'' performed much worse. This may suggest that the longer commands are easier for Network~2 to obtain the contexts.

\begin{table}
\begin{tabular}{l | rrr}
& User A & User B & ave.\\
\hline
Network~1 & 60.0\% & 25.0\% & 42.5\% \\ 
Network~1 + Network~2& 65.0\% & 65.0\% & 65.0\%  \\
GT & 90.0\% & 90.0\%  & 90.0\%  \\
\hline
\end{tabular}
\caption{Speech-recognition success ratio in tests with an unchanged existing smart speaker (``GT'' means Mel-scale encoded/decoded from the original voice data, regarded as the ground truth of the training).}
\label{fig:table}
\end{table}

\section{End-to-End Evaluation and Observations}

The real end-to-end silent voice to audio conversion was examined. In this case, a user is asked to mouth a speech command without actually emitting a sound, and the oral cavity movement is record by an ultrasonic imaging probe. The obtained image sequence is subsequently translated to a voice by the proposed system. 

We asked the participants to speak as silently as possible (not to vibrate their vocal cords) and to try to speak as similar as possible when they speak with sound. However, we did not ask them to hold their breaths. Consequently, a case occurred where a small leaked sound was audible. To clarify this situation, we measured the level of the emitted sound level from the participant by following the evaluation method of SilentVoice~\cite{Fukumoto:2018}. We used a noise meter that has the same specifications (min range $30\:dB$, $1.5\:dB$ error) placed $30\:cm$ away, in a room with a background noise level of $31.0\:dB(A)$. The mean of the peak sound level of 20 measurements (typical Alexa speech commands) was $37.14\:dB(A)$, which is lower than that of soft whispering.

Figure~\ref{fig:comparison} shows the comparison of the generated sound when a user is emitting a voice and when a user is not emitting a voice (please refer to the supplemental video for the actual generated sounds).

Initially, the result was unsatisfactory. We first expected that the movement in the oral cavity without emitting a voice (Figure~\ref{fig:comparison} (middle)) is the same as that when the user actually emits a voice (Figure~\ref{fig:comparison} (left)); however, a subtle difference was found between them, and the sound quality generated by the image-without-voice was not as good as that generated by the image-with-voice.

However, the following interesting phenomenon was observed. As the user can also listen to the generated sound generated by the image-without-voice, the user attempted to change the mouth movement to obtain a slightly better result. After several trials, the quality of generated sound improved. In this case, it is considered that the users improved their own skills in silent voicing.

\section{Discussions}

\begin{figure}
\includegraphics[width=0.4\textwidth]{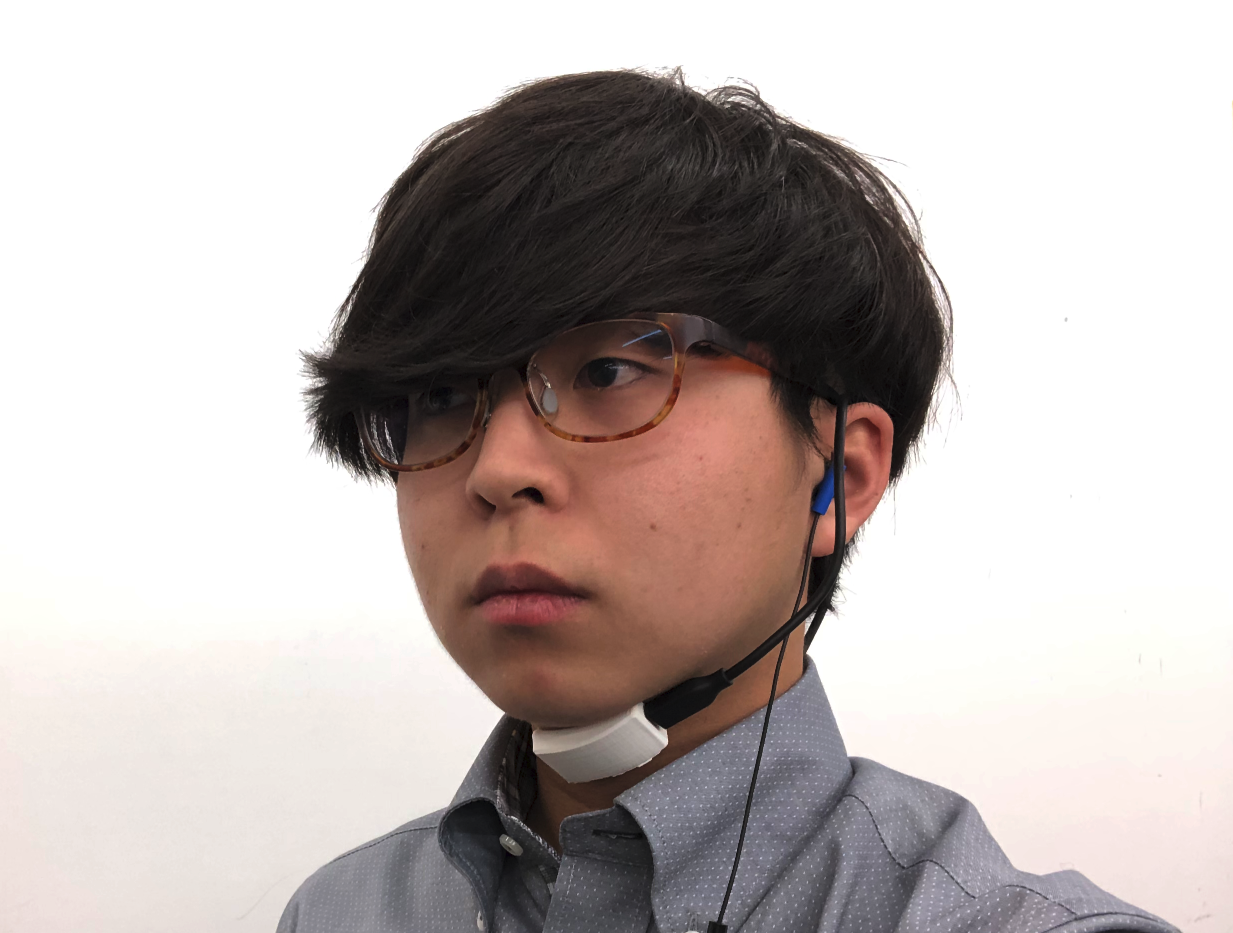}
\caption{Future Image: possible configuration of an ultrasonic probe and an open earphone}
\label{fig:future}
  \vspace{-0.4cm}
\end{figure}


\subsection{Incremental Voice Generation}
In our current design, the obtained ultrasonic-image sequences were converted to a voice at the granularity of the speech command (approximately 3.6 s). This is because Network~2 uses a fixed-length voice-representation sequence. However, based on the observation of the users' practice, it should be better to generate sounds incrementally, such that the user can have a tighter feedback loop for learning oral movements for generating a better voice.

\subsection{Continuous Ultrasonic-Wave Emission into the Body}

The effect on human organs when ultrasonic waves are emitted continuously into the body is unknown. However, we may be able to combine a simple triggering mechanism to start and stop the emission of ultrasonic waves. For example, the combination of an accelerometer and a microphone in the device can detect jaw movements for starting the (silent) voice command without actually emitting a voice.

\subsection{Cure for Vocal Cord Disabilities}
We also expect people with damaged vocal cords to use our research. As described in the Human--AI integration section, people will be able to learn how to correctly control their mouth and tongue to generate sound, even though their vocal cords do not work.

\subsection{Combining with Other Modalities}

Finally, it should be mentioned that this research is not intended to exclude other modalities. Combining the information of EMG, accelerometers, and NAM microphones may improve the quality of speech recognition. Investigating the combination of these modalities is subject of future research.

\subsection{Human--AI integration}

\begin{figure}
\centerline{}
\includegraphics[width=0.4\textwidth]{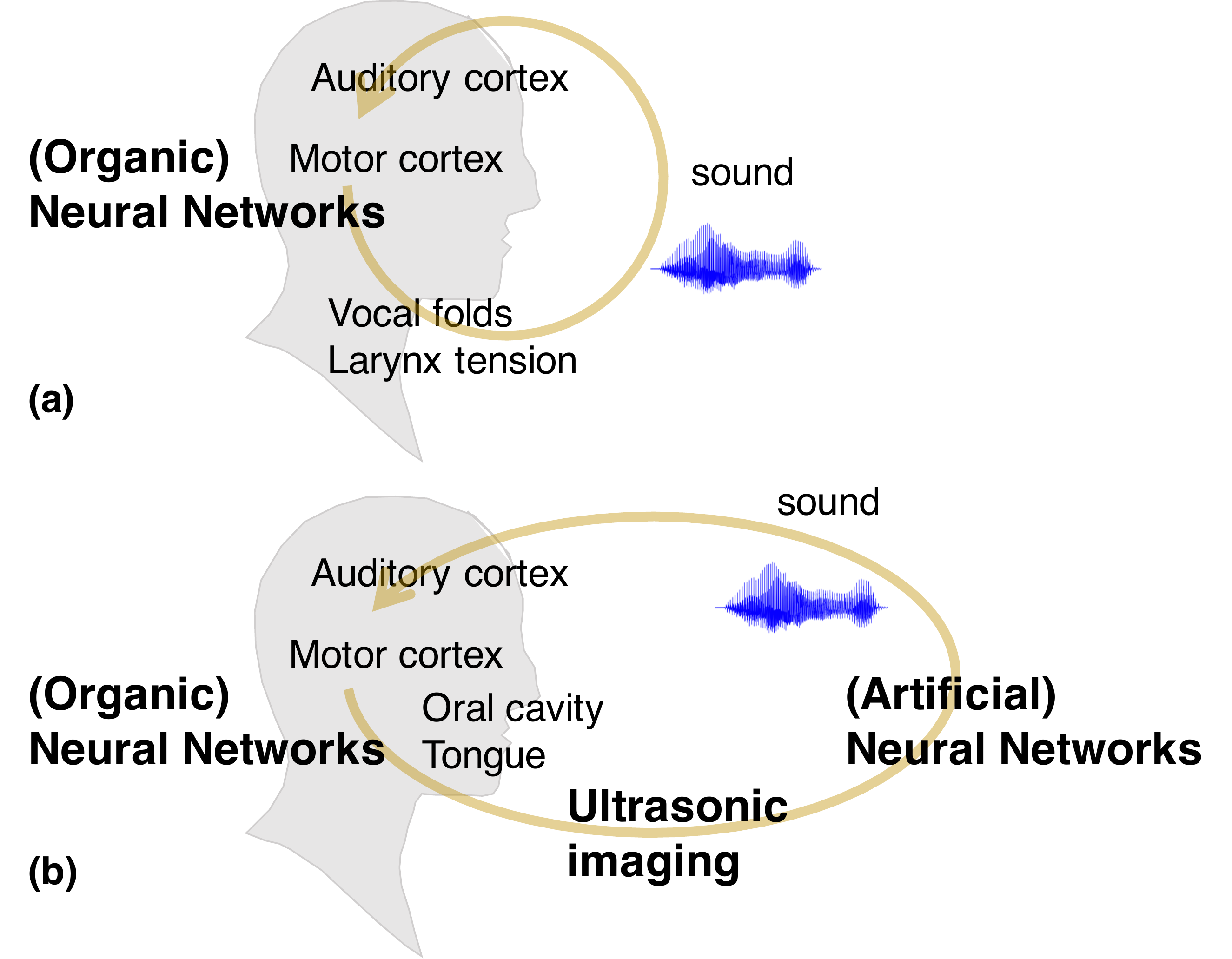}

\caption{Human--AI integration: artificial neural networks and organic neural networks form a feedback loop for obtaining better results.}
\label{fig:loop} 
\vspace{-0.4cm}
\end{figure}

The above-described observation suggests that an interesting relationship exists between humans and AI. Rather than considering AI as an autonomous or separated entity, we may be able to regard AI as a part of humans. Hence, even when the initial performance of the (artificial) neural networks is not perfect, a user can gradually learn and improve their performance.

This is similar to how people learn fundamental skills. When we learn utterance, the coordination of the motor cortex that drives the oral cavity, tongue, and vocal folds, and the auditory cortex forms a tight loop to obtain better speech performance (Figure~\ref{fig:loop} (a)). By extending this loop, organic neural networks (e.g., our brain) and artificial neural networks may also form a tight feedback loop (Figure~\ref{fig:loop} (b)). We name this formation, ``human--AI integration'' rather than human--AI interaction.

In this regard, we refer to the research of Glove Talk II from 1995 a pioneering work on human--AI integration~\cite{Fels:1995:GAG:223904.223966}. In this work, the user learned to control a voice synthesizer with hand gestures. Three simple neural networks were used, and the user (who was a pianist) required more than 100 h to generate an audible voice. A combination of better neural networks and a learner could reduce this learning time.

\section{Conclusion}
A method of silent-voice interaction with ultrasonic imaging was proposed. Two neural networks were used in sequence to convert a mouthed ``utterance'' of a user without a voice to a sound (voice), and could be used to operate the existing voice-controllable devices such as smart speakers. 

Following this result, we envision that the future form-factor for the wearable computer would be a combination of an attachable ultrasonic imaging probe to the underside of the jaw, with a bone conductive earphone or an open-air earphone (Figure~\ref{fig:future}). With this configuration, a user can always invoke a voice-controllable assistant without emitting a voice and obtain responses.

\bibliographystyle{ACM-Reference-Format}
\balance
\bibliography{main}

\end{document}